# Solutions of the bi-confluent Heun equation in terms of the Hermite functions

T.A. Ishkhanyan[1,2] and A.M. Ishkhanyan[1,3,4]
[1]Institute for Physical Research, NAS of Armenia, 0203 Ashtarak, Armenia
[2]Moscow Institute of Physics and Technology, Dolgoprudny, 141700 Russia
[3]Armenian State Pedagogical University, Yerevan 0010, Armenia
[4]Institute of Physics and Technology, National Research Tomsk Polytechnic University, Tomsk 634050, Russia

We construct an expansion of the solutions of the bi-confluent Heun equation in terms of the Hermite functions. The series is governed by a three-term recurrence relation between successive coefficients of the expansion. We examine the restrictions that are imposed on the involved parameters in order that the series terminates thus resulting in closed-form finite-sum solutions of the bi-confluent Heun equation. A physical application of the closed-form solutions is discussed. We present the five six-parametric potentials for which the general solution of the one-dimensional Schrödinger equation is written in terms of the bi-confluent Heun functions and further identify a particular conditionally integrable potential for which the involved bi-confluent Heun function admits a four-term finite-sum expansion in terms of the Hermite functions. This is an infinite well defined on a half-axis. We present the explicit solution of the one-dimensional Schrödinger equation for this potential and discuss the bound states supported by the potential. We derive the exact equation for the energy spectrum and construct an accurate approximation for the bound-state energy levels.

## 1. Introduction

The bi-confluent Heun equation is widely encountered in contemporary physics and mathematics research [1-5]. For example, in nuclear and atomic physics this equation frequently appears in studying the motion of quantum particles in one-, two- or three-dimensional confinement potentials [1]. The double-well quartic and sextic anharmonic oscillator potentials and the special class of singular confinement potentials consisting of a combination of Coulomb, linear and harmonic potentials are well-known examples of this class of potentials [6,7]. The recent examples include the inverse square root potential [8] and its conditionally exactly integrable generalization [9], applications to quantum chemistry [10], quantum dots [11], and quantum two-state systems [12].

Due to its wide appearance in theoretical physics, mathematical properties of the bi-confluent Heun equation have been studied by many authors (see, e.g., [1-3,13-26]). In particular, the power-series solutions near the regular singularity at the origin and in the neighborhood of the irregular singularity at the infinity [17,18], the continued fraction technique [19] and the Hill determinant approach [20] for a class of confinement potentials have been discussed in detail. Among the recent developments one may mention the following results. In [21,22] relations between the linear equations of the (deformed) Heun class and the six Painlevé nonlinear equations have been established via an anti-quantization procedure. In [23] the factorization of the confluent Heun equations is re-examined. In [24] $k$-summability is used to obtain new integral formulas for the solutions near the infinity and in [25] integral representations for a fundamental system of solutions to the bi-confluent Heun equation are derived using the properties of the Meijer $G$-functions. Integral equations for special functions of the Heun class are discussed in [26].

However, despite the large number of articles treating the bi-confluent Heun equation, the theory of this equation certainly needs further development. A particular observation appropriate in this instance is that the solutions are mostly constructed through power-series expansions and, as a consequence, the applications are mostly based on the polynomial reductions. However, the recent developments such as the exact solution of the Schrödinger equation for the inverse square root potential in terms of the Hermite functions of non-integer order [8], which are not polynomials (nor quasi-polynomials), suggests that the extension to the non-polynomial cases is an interesting and important challenge, as stated in [10]. Presumably, some useful properties of the solutions of the bi-confluent Heun equation may be revealed if expanding the solutions in terms of more advanced mathematical functions rather than powers.

In the present paper we make a step in this direction by constructing an expansion of the solutions of the bi-confluent Heun equation in terms of the Hermite functions. Being inspired by the two-term Hermite-function solutions constructed in [8,9], as expansion functions we apply a set of non-integer order Hermite functions of a shifted and scaled argument. As a result, we derive an expansion governed by a three-term recurrence relation between the successive coefficients of the expansion. Notably, under some restrictions imposed on the involved parameters the constructed series is terminated and thus closed-form finite-sum solutions of the bi-confluent Heun equation in terms of the Hermite functions are derived. This is a main result of the present paper.

Another result is the general solution of the one-dimensional stationary Schrödinger equation for a bi-confluent-Heun potential in terms of the Hermite functions that are not reduced to quasi-polynomials. The treatment is based on the constructed expansion and applies the described termination technique. This result demonstrates that the expansions of the Heun functions in terms of advanced special functions rather than simple powers suggest a quite productive approach. We note that a similar message follows from the recently reported results for a single-confluent-Heun [27] and a general-Heun [28] potentials.

As an application of the constructed solutions we consider the bi-confluent Heun potentials for the one-dimensional stationary Schrödinger equation [4,5,29]. The results indicate that there presumably exists an infinite number of conditionally exactly solvable potentials the solution for which is written as a linear combination of a finite number of Hermite functions of the above-mentioned type. The first potentials of the list are the harmonic oscillator potential [30] and the two Stillinger potentials [31] the solution for which involves just one Hermite function. The next come the first Exton potential [32] (involving the inverse square root potential [8] and its conditionally integrable generalization [9]) and the second Exton potential [32]. Te general solution for the latter two Exton potentials involves fundamental solutions each of which presents an irreducible combination of two Hermite functions. To give a representative example of a higher order expansion involving more terms, we here consider the case involving *four* Hermite functions. The corresponding potential is an infinite well defined on a half-axis. We present the four-term explicit solution of the Schrödinger equation for this potential and derive the exact energy-spectrum equation for the bound states that vanish both in the origin and at the infinity. Finally, we construct a highly accurate approximation for the bound-state energy levels.

**2. The expansion**

The bi-confluent Heun equation is a second-order ordinary linear differential equation, which has one regular singularity and an irregular singularity of rank 2 [1-3]. Conventionally, the regular singularity is put in the origin and the irregular one is located at the infinity. This is a confluent form of the general Heun equation derived via coalescence of its two finite regular singularities with the one located at the infinity [1-3].

Though the bi-confluent Heun equation involves only four irreducible parameters [1-3], for the sake of simplicity and generality we adopt here the following five-parametric form of this equation:

$$\frac{d^2u}{dz^2}+\left(\frac{\gamma}{z}+\delta+\varepsilon z\right)\frac{du}{dz}+\frac{\alpha z-q}{z}u=0, \qquad (1)$$

where $\gamma,\delta,\varepsilon,\alpha,q$ are arbitrary complex parameters. The advantage of this form is that the different canonical forms of the bi-confluent Heun equation as well as its limiting cases applied in the standard reference literature [1-3] and in numerous applications [4-26] are readily derived from this form by simple specifications of the involved parameters.

Based on the experience gained in solving the Schrödinger equation for the three-parametric inverse square root potential [8] and its four-parametric conditionally exactly solvable generalization [9], we consider the following expansion of the solution of the bi-confluent Heun equation (1) in terms of the Hermite functions of a shifted and scaled argument:

$$u = \sum_n c_n u_n, \quad u_n = H_{\alpha_0 + n}(s_0(z + z_0)), \tag{2}$$

where $\alpha_0, s_0$ and $z_0$ are complex constants to be defined afterwards. We note that in general the index parameter $\alpha_0$ is not an integer so that the expansion functions are not reduced to quasi-polynomials.

The Hermite functions satisfy the following second-order linear differential equation:

$$\frac{d^2 u_n}{dz^2} - 2s_0^2(z + z_0)\frac{du_n}{dz} + 2s_0^2 \alpha_n u_n = 0, \quad \alpha_n = \alpha_0 + n. \tag{3}$$

Substituting equations (2) and (3) into equation (1) and multiplying the result by $z$ we get

$$\sum_n c_n \left[ \left( \gamma + z(\delta + \varepsilon z) + 2s_0^2 z(z + z_0) \right) u_n' + \left( \alpha z - q - 2s_0^2 \alpha_n z \right) u_n \right] = 0. \tag{4}$$

This equation is considerably simplified if we put $2s_0^2 = -\varepsilon$ and $2s_0^2 z_0 = -\delta$, that is

$$s_0 = \pm\sqrt{-\varepsilon/2}, \quad z_0 = \delta/\varepsilon. \tag{5}$$

This choice cancels the terms proportional to $zu_n'$ and $z^2 u_n'$ so that using the recurrence identities

$$u_n' = 2s_0 \alpha_n u_{n-1} \tag{6}$$

and 
$$s_0(z + z_0)u_n = \alpha_n u_{n-1} + u_{n+1}/2, \tag{7}$$

we straightforwardly arrive at a three-term recurrence relation for coefficients $c_n$:

$$R_n c_n + Q_{n-1} c_{n-1} + P_{n-2} c_{n-2} = 0 \tag{8}$$

with 
$$R_n = \frac{\sqrt{2}}{\sqrt{-\varepsilon}}(\alpha_0 + n)\left(\alpha + (\alpha_0 + n - \gamma)\varepsilon\right), \tag{9}$$

$$Q_n = \mp\frac{\alpha\delta + (q + (\alpha_0 + n)\delta)\varepsilon}{\varepsilon}, \tag{10}$$

$$P_n = \frac{\alpha + (\alpha_0 + n)\varepsilon}{\sqrt{-2\varepsilon}}, \tag{11}$$

where the signs $\mp$ in the equation for $Q_n$ refer to the choices $s_0 = \pm\sqrt{-\varepsilon/2}$, respectively.

For the left-hand side termination of the series at $n = 0$, applying the initial conditions $c_{-2} = c_{-1} = 0$, $c_0 \neq 0$, we get $R_0 = 0$. This condition is satisfied if $\alpha_0 = 0$ or $\alpha_0 = \gamma - \alpha/\varepsilon$. The first choice $\alpha_0 = 0$ leads to the known polynomial solutions [1], hence, we discuss the second choice $\alpha_0 = \gamma - \alpha/\varepsilon$ which is applicable for non-zero $\varepsilon$. The summation index $n$ in expansion (2) then runs from zero to infinity and thus the final expansion is written as

$$u = \sum_{n=0}^{\infty} c_n H_{n+\gamma-\alpha/\varepsilon}\left(\pm\sqrt{-\varepsilon/2}\,(z+\delta/\varepsilon)\right). \tag{12}$$

We note that by choosing here different signs for the argument of the involved Hermite functions we get in general different independent solutions of the bi-confluent Heun equation (1) under consideration. Hence, by taking a linear combination, with arbitrary constant coefficients, of the two expansions corresponding to the plus and minus signs, we get an expansion for the general solution of equation (1).

The developed series may terminate from the right-hand side thus resulting in closed-form finite-sum solutions. This happens if two successive coefficients vanish for some $n = N = 0,1,2,...$, i.e., if $c_{N+1} = c_{N+2} = 0$ while $c_N \neq 0$. From equation $c_{N+2} = 0$ we find that the termination is possible if $P_N = 0$. This condition is satisfied if $\gamma = -N$. Since $\varepsilon$ is non-zero, the remaining equation $c_{N+1} = 0$ then presents a polynomial equation of the degree $N+1$ for the accessory parameter $q$ (we refer to this equation as $q$-equation), which defines, in general, $N+1$ values of $q$ for which the termination of the series occurs. To be specific, here are these equations for $N = 0, 1$ and 2:

$$\gamma = 0: \quad q = 0. \tag{13}$$

$$\gamma = -1: \quad q^2 - \delta q + \alpha = 0. \tag{14}$$

$$\gamma = -2: \quad q^3 - 3\delta q^2 + 2(\delta^2 + \varepsilon + 2\alpha)q - 4\alpha\delta = 0. \tag{15}$$

There are many physical situations when these equations are satisfied for a particular problem at hand. To demonstrate this, we now apply the developed series to the one-dimensional stationary Schrödinger equation.

## 3. Bi-confluent Heun potentials for the stationary Schrödinger equation

There are five six-parametric potentials for which the general solution of the one-dimensional stationary Schrödinger equation is written in terms of the bi-confluent Heun functions [4,5,29]. We present here the solution for these potentials applying the general approach proposed in [29] (see also [33]). The derivation lines are as follows.

The one-dimensional stationary Schrödinger equation for a particle of mass $m$ and energy $E$ in a potential $V(x)$ is

$$\frac{d^2\psi}{dx^2} + \frac{2m}{\hbar^2}(E - V(x))\psi = 0. \tag{16}$$

Applying the transformation of the independent variable $z = z(x)$, this equation is rewritten for the new argument $z$ as

$$\frac{d^2\psi}{dz^2} + \frac{\rho_z}{\rho}\frac{d\psi}{dz} + \frac{2m}{\hbar^2}\frac{E - V(z)}{\rho^2}\psi = 0, \tag{17}$$

where $\rho = dz/dx$. The further transformation of the dependent variable $\psi = \varphi(z) u(z)$ reduces this equation to the following equation for the new dependent variable $u(z)$:

$$\frac{d^2u}{dz^2} + \left(2\frac{\varphi_z}{\varphi} + \frac{\rho_z}{\rho}\right)\frac{du}{dz} + \left(\frac{\varphi_{zz}}{\varphi} + \frac{\rho_z}{\rho}\frac{\varphi_z}{\varphi} + \frac{2m}{\hbar^2}\frac{E - V(z)}{\rho^2}\right)u = 0. \tag{18}$$

This equation is the bi-confluent Heun equation (1) if

$$2\frac{\varphi_z}{\varphi} + \frac{\rho_z}{\rho} = \frac{\gamma}{z} + \delta + \varepsilon z \tag{19}$$

and

$$\frac{\varphi_{zz}}{\varphi} + \frac{\rho_z}{\rho}\frac{\varphi_z}{\varphi} + \frac{2m}{\hbar^2}\frac{E - V(z)}{\rho^2} = \frac{\alpha z - q}{z}. \tag{20}$$

Since the point $z = 0$ is the only singularity of the bi-confluent Heun equation located in the finite part of the complex $z$-plane, according to the approach of [29], in order to identify the potentials that are proportional to an energy-independent parameter and have a shape that is independent of both energy and that parameter, we search for the solutions of equations (19) and (21) applying the transformation

$$\rho = \frac{dz}{dx} = \frac{z^{m_1}}{\sigma} \tag{21}$$

with an integer or half-integer $m_1$ and arbitrary $\sigma$. Resolving equation (19), we then have

$$\varphi = z^{\alpha_0} e^{\alpha_1 z + \alpha_2 z^2}, \tag{22}$$

where the constants $\alpha_{0,1,2}$ are defined through $m_1$, $\sigma$ and the parameters involved in the target bi-confluent Heun equation (1). Substituting equations (21) and (22) into the remaining equation (20) and multiplying further the equation by $z^2$, we note that the last term of the obtained equation, that is $z^2(E - V(z))/\rho^2$, is a polynomial in $z$ of at most fourth degree. If the energy $E$ adopts arbitrary values and the potential is energy-independent, this is possible only if the two summands of this term, the one proportional to $E$ and the other proportional to $V(z)$, independently of each other, are polynomials of at most fourth degree [29]:

$$\frac{z^2}{\rho^2} = r_0 + r_1 z + r_2 z^2 + r_3 z^3 + r_4 z^4, \tag{23}$$

$$V(z)\frac{z^2}{\rho^2} = v_0 + v_1 z + v_2 z^2 + v_3 z^3 + v_4 z^4. \tag{24}$$

Equation (23) together with equation (21) leads to 5 admissible sets of integer or half-integer $m_1$ defined by the inequalities $0 \leq 2 - 2m_1 \leq 4$. Thus, $m_1 = -1, -1/2, 0, 1/2, 1$. With these $m_1$, equation (24) defines five independent six-parametric bi-confluent Heun potentials first identified by Lemieux and Bose [4]. The potentials are listed in Table 1, where for convenience both $z$- and $x$-representations of these potentials are presented since the $z$-representation is in some cases more useful for practical calculations while from the physical point of view just the $x$-representation matters. Note that the parameters $x_0$ and $\sigma$ are not shown in the table; to get the full representation for the potential and the corresponding coordinate transformation, one should everywhere make the replacement $x \to (x - x_0)/\sigma$.

| $m_1$ | Potential $V(z)$ | Coordinate transformation | Explicit potential $V(x)$ |
|---|---|---|---|
| $-1$ | $V_0' + \frac{V_1'}{z} + \frac{V_2'}{z^2} + \frac{V_3'}{z^3} + \frac{V_4'}{z^4}$ | $z = \sqrt{2x}$ | $V_0 + \frac{V_1}{\sqrt{x}} + \frac{V_2}{x} + \frac{V_3}{x^{3/2}} + \frac{V_4}{x^2}$ |
| $-1/2$ | $V_0' + V_1' z + \frac{V_2'}{z} + \frac{V_3'}{z^2} + \frac{V_4'}{z^3}$ | $z = (3x/2)^{2/3}$ | $V_0 + V_1 x^{2/3} + \frac{V_2}{x^{2/3}} + \frac{V_3}{x^{4/3}} + \frac{V_4}{x^2}$ |
| $0$ | $V_0' + V_1' z + V_2' z^2 + \frac{V_3'}{z} + \frac{V_4'}{z^2}$ | $z = x$ | $V_0 + V_1 x + V_2 x^2 + \frac{V_3}{x} + \frac{V_4}{x^2}$ |
| $1/2$ | $V_0' + V_1' z + V_2' z^2 + V_3' z^3 + \frac{V_4'}{z}$ | $z = x^2/4$ | $V_0 + V_1 x^2 + V_2 x^4 + V_3 x^6 + \frac{V_4}{x^2}$ |
| $1$ | $V_0' + V_1' z + V_2' z^2 + V_3' z^3 + V_4' z^4$ | $z = e^x$ | $V_0 + V_1 e^x + V_2 e^{2x} + V_3 e^{3x} + V_4 e^{4x}$ |

Table 1. Five six-parametric bi-confluent Heun potentials ($V_{0,1,2,3,4}$ are arbitrary constants) together with the corresponding coordinate transformation $z = z(x)$.

For these potentials together with corresponding $m_1$, collecting the coefficients at powers of $z$ in equations (19) and (20), we get eight equations which are linear for the parameters $\gamma, \delta, \varepsilon, \alpha, q$ of the bi-confluent Heun function $u(z)$ as well as for the parameter $\alpha_1$ of the pre-factor $\varphi(z)$, and are quadratic for the parameters $\alpha_0$ and $\alpha_2$. The solution of the stationary Schrödinger equation (16) for the presented potentials is thus explicitly written in terms of the bi-confluent Heun function as

$$\psi = z^{\alpha_0} e^{\alpha_1 z + \alpha_2 z^2} H_B(\gamma, \delta, \varepsilon; \alpha, q; z) \tag{25}$$

with the involved parameters being given by the equations

$$\gamma = 2\alpha_0 + m_1, \quad \delta = 2\alpha_1, \quad \varepsilon = 4\alpha_2, \tag{26}$$

$$\alpha = \alpha_1^2 + 2\alpha_2(2\alpha_0 + m_1 + 1) + 2m(Er_2 - v_2)/\hbar^2, \tag{27}$$

$$q = -\alpha_1(2\alpha_0 + m_1) - 2m(Er_1 - v_1)/\hbar^2 \tag{28}$$

and

$$\alpha_0(\alpha_0 + m_1 - 1) + 2m(Er_0 - v_0)/\hbar^2 = 0, \tag{29}$$

$$\alpha_1 \alpha_2 + m(Er_3 - v_3)/(2\hbar^2) = 0, \tag{30}$$

$$\alpha_2^2 + m(Er_4 - v_4)/(2\hbar^2) = 0. \tag{31}$$

This fulfills the development. Starting from a particular potential of Table 1, one calculates, through equations (26)-(31), the parameters $\gamma, \delta, \varepsilon, \alpha, q$ of the bi-confluent Heun function $H_B$ and the parameters $\alpha_{0,1,2}$ of the pre-factor $\varphi(z)$. We note that for a given potential with corresponding $m_1$ the auxiliary parameters $r_{0,1,2,3,4}$ and $v_{0,1,2,3,4}$ involved in these equations are readily calculated through the definitions (23) and (24). A last remark concerns the factor $\sigma$ involved in equation (21) and the constant $x_0$ which comes out from integration of equation (21). It is immediately seen from the form of the presented potentials that, since $V_{0,1,2,3,4}$ are arbitrary, without loss of the generality one may put $\sigma = 1$. As regards $x_0$, depending on the particular physical problem at hand, e.g., if a possible non-Hermitian extension is considered, this constant standing for the space origin may be chosen complex.

Thus, we have presented the solution of the Schrödinger equation for the five bi-confluent-Heun Lemieux-Bose potentials. If we now apply the expansion (12) of the bi-confluent Heun function in terms of the Hermite functions, we may check if the conditions (13)-(15) for termination of the series for some $n = N$ are fulfilled for particular choices of the parameters involved in the potential. If yes, then we arrive at the solution of the Schrödinger equation (1) for that particular potential written as a sum of a finite number of the Hermite functions:

$$\psi(z) = z^{\alpha_0} e^{\alpha_1 z + \alpha_2 z^2} \sum_{n=0}^{N} c_n \, H_{(\gamma-\alpha/\varepsilon)+n}\left(\pm\sqrt{-\frac{\varepsilon}{2}}\left(z + \frac{\delta}{\varepsilon}\right)\right). \quad (32)$$

The inspection shows that there are several cases for which the result of the test is positive. For instance, the termination conditions can be satisfied for $m_1 = -1/2$. The results are as follows. For $N = 0$ we get the exactly solvable general harmonic oscillator potential [30] and the two conditionally exactly solvable potentials by Stillinger [31]. The solution of the Schrödinger equation for these potentials involves only one Hermite function. For $N = 1$ we get the inverse square root potential [8] and its conditionally solvable generalization [9]. For these potentials the solution involves two Hermite functions. We note that the pointed generalized conditionally exactly solvable potential was first identified by Exton (Eq. (21) of [32]). We refer to this potential as the first conditionally integrable Exton potential. It is worth mentioning that it involves as particular cases also the two super-symmetric partner potentials treated by Lopez-Ortega [34] (note that these SUSY partner potentials can also be derived by considering the inverse square root potential as a super-potential). Apart from this first potential by Exton, the case $N = 1$ yields also the second conditionally integrable potential proposed by Exton as well (Eq. (22) of [32]). It is understood that the cases $N = 2, 3, ...$ may result in an infinite sequence of conditionally exactly solvable potentials. As a representative example, we present a higher-order termination with $N = 3$, i.e. a potential for which the solution of the Schrödinger equation is a linear combination of *four* Hermite functions.

## 4. A bi-confluent Heun potential solvable in terms of the Hermite functions

Let the series (12) terminate at $N = 3$, that is $c_3 \neq 0$ and $c_4 = c_5 = 0$. For this to be the case should be $\gamma = -3$ and the accessory parameter $q$ should satisfy the equation

$$q^4 - 6\delta q^3 + (11\delta^2 + 10\varepsilon + 10\alpha)q^2 - 6\delta(\delta^2 + 3\varepsilon + 5\alpha)q + 9\alpha(2\delta^2 + 2\varepsilon + \alpha) = 0. \quad (33)$$

Let $m_1 = -1/2$. By checking the parameters $\gamma, \delta, \varepsilon, \alpha, q$ obeying equations (26)-(31) for fulfillment of equation (33), we readily get that this is the case for the potential

$$V(x) = \frac{55\hbar^2}{72 m\, x^2} + \frac{V_2}{x^{2/3}} + V_0 + \frac{9 m V_2^2}{8\hbar^2} x^{2/3}. \quad (34)$$

Since one can everywhere make the replacement $x \to x - x_0$ with an arbitrary $x_0$ standing for the coordinate origin, this is in general a three-parametric potential. We note that, since this potential involves a fixed parameter ( $55\hbar^2/(72m)$ ) and the strengths of the terms proportional to $x^{-2/3}$ and $x^{2/3}$ do not vary independently, this is a conditionally integrable potential. The shape of the potential is shown in Fig. 1. We note that a reason for the potential to be of interest is that owing to the involved centrifugal-barrier term it models, to a certain extent, the one-dimensional reduction of the three-dimensional Schrödinger problem for the central fractional-power singular potential involving $r^{-2/3}$ and $r^{2/3}$ terms used in the past in particle physics phenomenology [35,36].

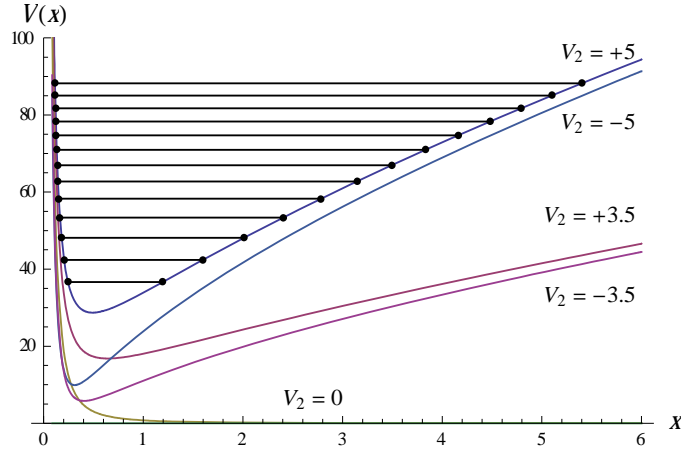

Fig.1 Potential (34) for $V_0 = 0$ and $V_2 = 0, \pm 3.5, \pm 5$ ($m = \hbar = 1$).

According to the expansion (32), the fundamental solutions $\psi_F$ of the Schrödinger equation (16) for this potential are written as a sum of four Hermite functions:

$$\psi_F(z) = z^{-5/4} e^{-y^2/2} \left( c_0 H_{a-3}(y) + c_1 H_{a-2}(y) + c_2 H_{a-1}(y) + c_3 H_a(y) \right), \tag{35}$$

where
$$y = \sqrt{-2\alpha_2} \left( \left( \frac{3x}{2} \right)^{2/3} + \frac{\alpha_1}{2\alpha_2} \right). \tag{36}$$

The index parameter $a$ as well as the expansion coefficients $c_{0,1,2,3}$ are conveniently written through the pre-factor exponents $\alpha_1$ and $\alpha_2$ (which are explicitly calculated by equations (29) and (31), respectively) and an auxiliary parameter $s = \pm 1$. The result reads

$$a = 1 + s - \frac{\alpha_1^2}{4\alpha_2}, \tag{37}$$

and
$$c_0 = 8\sqrt{-2\alpha_2}\, a(a-1)(a-2), \quad c_1 = 12\alpha_1 a(a-1), \tag{38}$$

$$c_2 = 6\sqrt{-2\alpha_2}\,a(2a-3-s), \quad c_3 = \alpha_1(2a-1+s). \tag{39}$$

with
$$\alpha_1 = \left(\frac{2}{3}\right)^{2/3}\frac{V_3-E}{V_2 s}, \quad \alpha_2 = \left(\frac{3}{2}\right)^{2/3}\frac{mV_2 s}{2\hbar^2}. \tag{40}$$

We note that $s=+1$ and $s=-1$ produce linearly independent solutions. This is readily verified by checking the Wronskian of the two solutions. Hence, the linear combination of these fundamental solutions

$$\psi(z) = C_1\,\psi_F\big|_{s\to+1} + C_2\,\psi_F\big|_{s\to-1} \tag{41}$$

with arbitrary constant coefficients $C_{1,2}$ presents the general solution of the problem.

## 5. Bound states.

The bound states supported by potential (34) are derived by imposing the boundary conditions of vanishing the wave-function (41) in the origin and at the infinity: $\psi(0)=0$ and $\psi(\infty)=0$. The first condition presents a linear relation between the coefficients $C_{1,2}$:

$$C_1\psi_F(x=0)\big|_{s=+1} + C_2\psi_F(x=0)\big|_{s=-1} = 0, \tag{42}$$

while the second condition for the infinity reveals, after some algebra when passing to the large-argument asymptotes of the involved Hermite functions, that $C_1=0$ if $V_2>0$ and $C_2=0$ if $V_2<0$. It is then readily seen from equation (42) that for the bound states it holds

$$\psi_F(x=0)\big|_{s=-\mathrm{sign}(V_2)} = 0. \tag{43}$$

This is the exact equation for the energy spectrum and the bound-state wave functions are given by the first term of the general solution (41) if $V_2<0$ and by the second term if $V_2>0$. Combining the two cases, we thus have

$$\psi_B(x) = C_N\,\psi_F(x)\big|_{s=-\mathrm{sign}(V_2)}, \tag{44}$$

where $C_N$ is the normalization constant:

$$C_N = \sqrt{\int_0^{+\infty}\left(\psi_F\psi_F^*\right)\big|_{s=-\mathrm{sign}(V_2)}\,dx}, \tag{45}$$

It turns out that the spectrum equation is considerably simplified when separately considering positive and negative $V_2$ cases. Indeed, consider, for example, the case $V_2<0$ for which $s=+1$. Since the pre-factor $z^{-5/4}e^{-y^2/2}$ of the fundamental solution $\psi_F$ does not adopt zeros, equation (43) reduces to

$$\left(c_0 H_{a-3} + c_1 H_{a-2} + c_2 H_{a-1} + c_3 H_a\right)\Big|_{y=y_0} = 0, \tag{46}$$

with
$$y_0 \equiv y\big|_{x=0,\,s=1} = -\frac{\alpha_1}{\sqrt{-2\alpha_2}}\bigg|_{s=1} = -\sqrt{2(a-2)}. \tag{47}$$

Using the recurrence relations between the involved Hermite functions, equation (46) can be rewritten in a two-term form:

$$\left(A_1 H_{a-1} + A_2 H_a\right)\big|_{y=y_0} = 0 \tag{48}$$

with
$$A_1 = 2(a-2)(c_1 y_0 + (a-1)c_2) + c_0\left(1-a+2y_0^2\right), \tag{49}$$

$$A_2 = c_0 y_0 + (a-2)c_1 - 2\left(2-3a+a^2\right)c_3. \tag{50}$$

Notably, the inspection reveals that for the parameters (37)-(40) with $s = +1$ the second of the coefficients $A_{1,2}$ identically vanishes for arbitrary parameters of the potential (34) (with the proviso $V_2 < 0$), i.e., $A_2 \equiv 0$. We then arrive at a remarkably simple eigenvalue equation for the energy spectrum:

$$H_{a-1}\left(-\sqrt{2(a-2)}\right)\big|_{s=1} = 0, \quad V_2 < 0. \tag{51}$$

Similarly, for a positive $V_2$ we get

$$H_{a-2}\left(-\sqrt{2a}\right)\big|_{s=1} = 0, \quad V_2 > 0. \tag{52}$$

It should be stressed that these are exact equations. The first energy levels for the potential parameters $V_0 = 0$ and $V_2 = +5$ ($m = \hbar = 1$) are shown in Fig. 1. The corresponding normalized bound-state wave functions are plotted in Fig. 2.

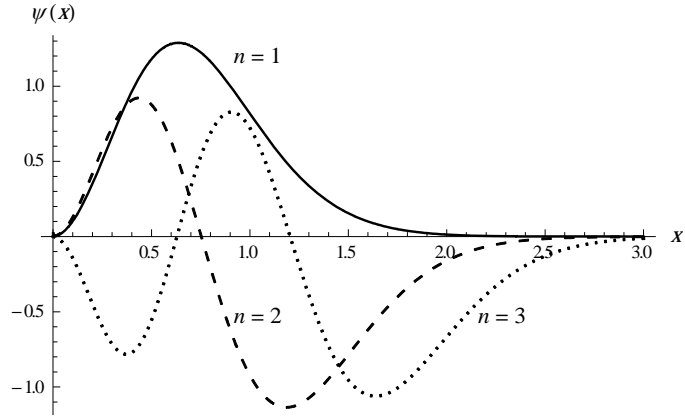

Fig. 2. The first three normalized wave functions for $V_0 = 0$, $V_2 = +5$ ($m = \hbar = 1$).

Consider now the approximate solution of the eigenvalue equations (51),(52). Note that the indexes and arguments of the Hermite functions $H_\nu(w)$ involved in these equations belong to the transition layer where $w \approx \pm\sqrt{2\nu}$. An appropriate approximation of the Hermite function for this region is [37]:

$$H_\nu(w) \propto 2^{\frac{1+\nu}{2}} e^{\frac{w^2-\nu+\nu\ln\nu}{2}} \left(1-\frac{w^2}{2\nu}\right)^{-1/4} \cos\left(\frac{\pi\nu}{2} - w\sqrt{\frac{\nu}{2}-\frac{w^2}{4}} - \frac{2\nu+1}{2}\arcsin\left(\frac{w}{\sqrt{2\nu}}\right)\right). \tag{53}$$

This is an accurate approximation that includes an appropriate factor for changing the amplitude and takes into account the uneven spacing of the zeros of the Hermite function. Importantly, the approximation is applicable to the whole admissible variation range $a > 2$ of the parameter $a$, which is the only parameter involved in equations (52) and (53). The approximation for the function $H_{a-2}(-\sqrt{2a})$ is shown in Fig. 3.

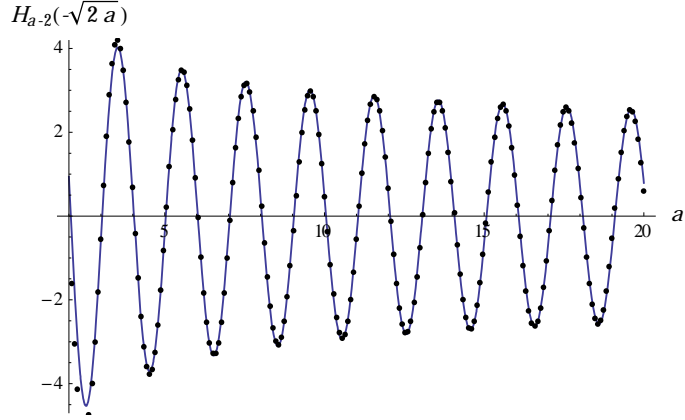

Fig. 3. Approximation (53) (solid line) for function $H_{a-2}\left(-\sqrt{2a}\right)$ (points).

Owing to the approximation (53) the spectrum equation is accurately approximated as

$$\frac{\pi\nu}{2} - w\sqrt{\frac{\nu}{2}-\frac{w^2}{4}} - \frac{2\nu+1}{2}\arcsin\left(\frac{w}{\sqrt{2\nu}}\right) = -\frac{\pi}{2} + \pi n, \quad n = 1, 2, \ldots. \tag{54}$$

The next step is now to express the parameters $\nu = a-1, a-2$ and $w = -\sqrt{2(a-2)}, -\sqrt{2a}$ through the energy $E$, using the definition (37) together with equations (38)-(40), and further make in equation (54) the substitution $E_n = V_3 + \left(9m|V_2|^3/(2\hbar^2)\right)^{1/2} f(n)$, which is readily guessed from the form of the derived expressions. The last step is then to expand $f(n)$ for

large $n$ into the series in terms of half-integer and integer powers of $n$. The resultant expansions for the eigenvalues up to the accuracy of the order of $O(1/n)$ are given as

$$E_n = V_3 + \sqrt{\frac{9m|V_2|^3}{2\hbar^2}}\left(\sqrt{n+1} - \frac{7}{8\sqrt{n+1}}\right) \quad \text{if} \quad V_2 < 0 \qquad (55)$$

and

$$E_n = V_3 + \sqrt{\frac{9m|V_2|^3}{2\hbar^2}}\left(\sqrt{n+1} + \frac{1}{64}\frac{1}{\sqrt{n+1}}\right) \quad \text{if} \quad V_2 > 0. \qquad (56)$$

These approximations are compared with the exact energy eigenvalues in Fig. 4. As seen, these are rather accurate results. Precisely, except for the first level, the relative error is less than $2.5 \times 10^{-3}$ for all $n \geq 2$ (this is shown in the inset of Fig. 4).

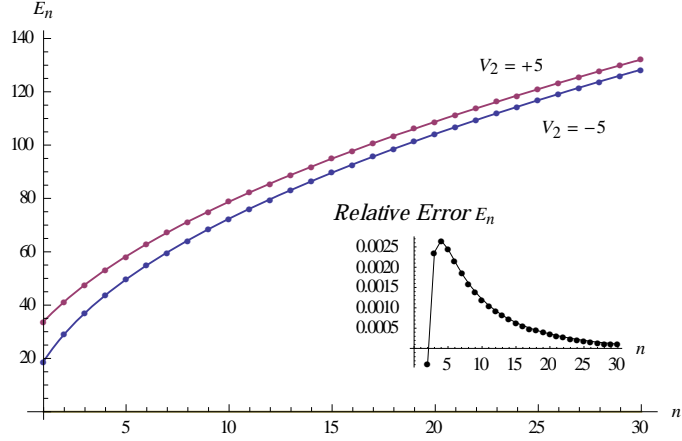

Fig. 4. Approximations (55),(56) (solid lines) versus exact energy levels (points) for $V_0 = 0$, $V_2 = \pm 5$ ($m = \hbar = 1$). The inset presents the relative error for $n > 1$.

## 6. Discussion

Thus, we have constructed an expansion of the general solution of the bi-confluent Heun equation in terms of the Hermite functions of a shifted and scaled argument. The expansion functions are in general of non-integer order so that they in general are not quasi-polynomials. The expansion applies for arbitrary sets of the involved parameters (with the proviso $\varepsilon \neq 0$). The coefficients of the expansion obey a three-term recurrence relation between the successive coefficients. We have shown that the constructed series may terminate thus resulting in closed-form solutions of the bi-confluent Heun equation involving a finite number of the Hermite functions. The restrictions imposed on the involved parameters in order that the series allow termination are: i) the characteristic (Frobenius) exponent $\mu = 1-\gamma$ of the regular singularity in the origin $z = 0$ should be a positive integer (this is achieved if $\gamma = -N$ with $N = 0,1,2,...$) and ii) the accessory parameter $q$ should obey a polynomial equation of the degree $N+1$ ($q$-equation). The resultant solution of the bi-confluent Hen equation then presents a linear combination of $N+1$ Hermite functions of a shifted and scaled argument presented above. This is a main result of the present paper.

We have presented the explicit $q$-equations for the accessory parameter for $N = 0,1,2$. Furthermore, when discussing a particular application to the one-dimensional Schrödinger equation, we have also presented the $q$-equation for $N = 3$. This is a representative example of higher-order termination resulting in a solution written as a linear combination of four Hermite functions.

As a representative application of the constructed solutions, we have discussed the reduction of the one-dimensional Schrödinger equation to the bi-confluent Heun equation. We have presented the five six-parametric Lemieux-Bose potentials and have constructed the solution of the problem for these potentials in terms of the bi-confluent Heun functions. Further, we have identified a sub-potential for which the involved parameters allow termination of the Hermite-function expansion of the involved bi-confluent Heun function at $N = 3$. This is a three-parametric conditionally integrable potential for which the general solution of the Schrödinger equation is written through fundamental solutions each of which presents a linear combination of four Hermite functions.

The derived potential is an infinite well defined on a half-axis. It involves a repulsive centrifugal-barrier core $\sim x^{-2}$ and fractional-power terms proportional to $x^{-2/3}$ and $x^{2/3}$. The strength of the first term is fixed and those for the latter two terms do not vary independently. This is why it is a conditionally integrable potential.

We have presented the four-term explicit solution of the Schrödinger equation for this potential in terms of the Hermite functions and have discussed the bound states supported by the potential. The exact energy-spectrum for the bound states that vanish both in the origin and at the infinity is defined by the zeros of a Hermite function of a non-integer order. However, the parameters of this Hermite function depend on the sign of the coefficient $V_2$ of the singular term proportional to $x^{-2/3}$ (the term $x^{2/3}$ is always repulsive). It is understood that this is because the case $V_2 = 0$ is exceptional in that then both fractional-power terms vanish so that in this case the potential is not a well (see Fig. 1).

Finally, we have considered the approximate solution of the eigenvalue equations for both negative and positive $V_2$. It is worth mentioning that the parameters of the Hermite functions $H_v(w)$ involved in these equations in both cases belong to the transition layer for which $w \approx \pm\sqrt{2v}$. Applying a specific asymptotic expansion applicable to the whole variation range of the parameters within this layer, we have derived highly accurate approximations for the bound-state energy levels written as a linear combination of half-integer powers of the quantum number $n+1$: $(n+1)^{1/2}$ and $(n+1)^{-1/2}$, for both negative and positive $V_2$.

Taking finally a general look at the presented results, we find that, apart from the very expansion that we have reported and the particular Schrödinger problem that we have treated, a general message of the present paper is that the expansions of the Heun functions in terms of advanced special functions rather than simple powers suggest a quite productive approach. In particular, the expansion of the bi-confluent Heun function in terms of the Hermite functions allows construction of closed-form finite-sum non-polynomial solutions.

Such solutions may be useful for treating many physical problems. Though we have convinced in this only by discussing a particular bi-confluent-Heun conditionally integrable potential for the one-dimensional Schrödinger equation, it is well understood that the conclusion is general and applies to other problems such as, e.g., the relativistic evolution governed by the Dirac or the Klein-Gordon equations [38,39]. Supporting this statement are the applications of the recent expansions of the solutions of the single-confluent Heun equation in terms of the Kummer confluent hypergeometric [40], the incomplete Beta and the Appell generalized hypergeometric [41] functions to the quantum two-state problem [42,43].

Expansions of the solutions of the Heun equations in terms of more advanced functions instead of simple powers have been initiated by Svartholm [44] and Erdélyi [45] who proposed series expansions of the general Heun functions in terms of the Gauss hypergeometric functions. This is a useful extension of the series technique applicable to many other differential equations including those of more general type such as the five *deformed* Heun equations which are the Heun equations with an additional *apparent* singularity [22]. As already stated above, a useful property suggested by these expansions is the possibility to derive finite-sum solutions by means of termination of the series. However, it should be noted that the infinite series themselves are also of notable interest as it is the case of the expansions of the general Heun functions in terms of the incomplete Beta functions applied to the surface plasmon-polariton problem [46]. The general conclusion is then that there is a pronounced need to explore the variety of all possible expansions of the Heun functions in terms of the functions of the hypergeometric class which currently form the most developed and most familiar set of special functions. Because of the enormous number of appearances of the Heun equations in contemporary classical and non-classical science one may envisage many important applications of these expansions.

**Acknowledgments**

This research has been conducted within the scope of the International Associated Laboratory IRMAS (CNRS-France & SCS-Armenia). The work has been supported by the Armenian State Committee of Science (SCS Grant No. 15T-1C323), Armenian National Science and Education Fund (ANSEF Grant No. PS-4558), and the project "Leading Russian Research Universities" (Grant No. FTI_24_2016 of the Tomsk Polytechnic University). T.A. Ishkhanyan acknowledges the support from SPIE through a 2017 Optics and Photonics Education Scholarship.